\documentclass[twocolumn,aps,prd,showpacs, nofootinbib]{revtex4}
\usepackage{graphicx}
\usepackage{amssymb}
\usepackage{amsmath}
\usepackage{bm}
\usepackage{times}\usepackage[caption=false]{subfig} 
\usepackage{subfig}

\newcommand{\msun}{M_\odot}
\newcommand{\be}{\begin{equation}}
\newcommand{\ee}{\end{equation}}
\newcommand{\bea}{\begin{eqnarray}}
\newcommand{\eea}{\end{eqnarray}}

\newcommand{\etal}{{\it et al.}}

\begin{document}

\title{Efficiency of nonspinning templates in gravitational wave searches for aligned-spin binary black holes}

\author{Hee-Suk Cho}
\email{chohs1439@pusan.ac.kr}
\affiliation{Korea Institute of Science and Technology Information, Daejeon 34141, Korea}

\date{\today}

\begin{abstract}

We study the efficiency of nonspinning waveform templates in gravitational wave searches for aligned-spin binary black holes (BBHs).
We use PhenomD, which is  the most recent phenomenological waveform model designed to generate the
full inspiral-merger-ringdown waveforms emitted from BBHs with the spins aligned with the orbital angular momentum.
Here, we treat the effect of aligned-spins with a single spin parameter $\chi$.
We  consider the BBH signals with moderately small spins in the range of $-0.4\leq \chi  \leq 0.4$.
Using nonspinning templates, we calculate fitting factors  of the aligned-spin signals  in a wide mass range up to $\sim 100 \msun$.
We find that the range in spin over which the nonspinning bank has fitting factors exceeding the threshold of 0.965 
for all signals in our mass range is very narrow, i.e.,  $-0.3\leq \chi \leq 0$.
The signals with negative spins can have higher fitting factors than those with positive spins.
If $\chi = 0.3$, only the highly asymmetric-mass signals can have the fitting factors exceeding the threshold,
while the fitting factors for all of the signals can be larger than the threshold  if $\chi = -0.3$. 
We demonstrate that the discrepancy between the regions of a positive and a negative spin is due to the physical boundary ($\eta \leq 0.25$) of
the template parameter space. In conclusion, we emphasize the necessity of an aligned-spin template bank in the current Advanced LIGO searches for aligned-spin BBHs.
We also show that the recovered mass parameters can be significantly biased  from the true parameters.

\end{abstract}


\maketitle

\section{Introduction}
Recently, two gravitational wave (GW) signals named as GW150914 and GW151226, were detected by the two LIGO detectors \cite{GW1,GW2},
and these observations indicate that future observing runs of the advanced detector network \cite{ALIGO,AVirgo,KAGRA} will yield more binary black hole (BBH) merger signals \cite{Aba10,Dom14,Abb16,Abb16b}.
Detailed analyses in the parameter estimation showed that both signals were emitted from merging BBHs \cite{GW1PE1,GW1PE2,GW2}.
The masses of the two binaries were found to be $\sim 65$ and $22 \msun$ for GW150914 and GW151226, respectively.  
In particular, the two components of GW150914 are the heaviest stellar mass BHs known to date.
On the other hand, the precession effects for both signals were poorly measured, while
the aligned-spins ($\chi$) were meaningfully constrained.
Although  we might expect high-spin BHs from the X-ray observations \cite{Nie16},
both binaries had small values of $\chi$. The $90 \%$ credible intervals in their parameter estimations were in the range of $-0.4 \leq \chi \leq 0.4$.

The waveforms emitted from BBHs have three phases: inspiral, merger, and ringdown (IMR),
and the IMR phases of stellar mass BBHs are likely to be captured in the sensitivity band of ground-based detectors.
In the search for BBHs, therefore, we have to use the full IMR waveforms as templates.
Over the past decade, two classes of IMR waveform models have been developed: effective-one-body models calibrated to numerical relativity simulations (EOBNR) and phenomenological   models.
Since EOBNR is formulated in the time domain
as a set of differential equations, generation of those waveforms are computationally much more expensive than generation of
frequency-domain waveforms.
Therefore, for the purpose of the GW data analysis,
P{\"u}rrer \cite{Pur14,Pur16} has recently built a Fourier-domain reduced order model that faithfully represents the original EOBNR model \cite{Tar12,Tar14}.
On the other hand, a series of the phenomenological models have been developed, and those were also constructed in the frequency domain.
The first phenomenological
model was PhenomA \cite{Aji07a,Aji08a,Aji08b} that was designed to model the IMR waveforms of nonspinning BBHs,
and this model was extended to an aligned-spin system in PhenomB \cite{Aji11b} by adding the effective spin parameter $\chi$.
The third model was PhenomC \cite{San10} that was also designed for aligned-spin BBHs, and extended to a 
precessing system in  PhenomP \cite{Han14}.
The most recent phenomenological model is PhenomD \cite{Kha16}.
This model is also designed for aligned-spin BBHs but
covers much wider ranges of mass (up to mass ratios of $1:18$) and spin (up to $|\chi|\sim 0.85$) than any other phenomenological models.
Recently, it has  been shown that PhenomD can perform very well for BBH searches, losing less than $1\%$ of the recoverable signal-to-noise ratio \cite{Kum16}.
Therefore, we use PhenomD for the waveforms of aligned-spin BBHs in this work\footnote{
The recent version of EOBNR reduced order model was also calibrated in wide parameter ranges up to mass ratios of $1:100$ and spins of $-1 \leq \chi_i\leq 0.99$ \cite{Pur16}.}.

We study the efficiency of nonspinning waveform templates in GW searches for aligned-spin BBH signals
by investigating the fitting factor.
The fitting factor is defined as the best-match between a normalized signal and a set of normalized templates \cite{Apo95}.
For the GW data analysis  purposes, the fitting factor is considered to evaluate the search efficiency.
Since the detection rate is proportional to $\rho^{1/3}$, 
a ${\rm FF}\simeq 0.97$ corresponds to a loss of detection rates of $\sim 10\%$.
Similar works have been carried by several authors in the past few years.
Using the post-Newtonian waveform model, Ajith \cite{Aji11a} calculated fitting factors  for several binaries with masses of $M \leq 20 \msun$.
He found that the spin value of the signal clearly  separated the population of binaries producing a poor fitting factor from those producing a high fitting factor  (see Fig. 11 therein).
Dal Canton \etal \cite{Dal14} also calculated fitting factors for BH-neutron star (NS) binaries with masses of $M \leq 18 \msun$, and
they also found a clear separation between  the two populations  (see Fig. 8 therein).
In the same paper, the authors showed that this behavior was due to the fact that the template parameter space is physically bounded as $\eta \leq 0.25$ (see Figs. 5 and 6 therein). 
A similar work has also been performed by Privitera \etal \cite{Pri14} for BBH systems with $10 \msun \leq M \leq 30 \msun$ and $m_1/m_2 \leq 4$
using the PhenomB waveforms \cite{Aji11b}.
They found that a nonspinning template bank  achieved
fitting factors exceeding 0.97  over a wide region of parameter space, spanning
roughly $-0.25 \leq \chi \leq 0.25$ over the entire mass range considered in their work (see Fig. 1 therein).
Recently, the work of \cite{Pri14} has been extended  to higher-mass systems $M\leq 50 \msun$ by Capano \etal \cite{Cap16}
using the EOBNR waveforms \cite{Tar14}.
On the other hand, several works have used precessing signals to test nonspinning and aligned-spin template banks \cite{Aji14,Har14,Dal15,Har16}.

In this work, we revisit the issues on the effectualness of nonspinning templates 
for aligned-spin BBH signals.
Although the template bank used for current  Advanced LIGO searches covers the binary masses up to $100 \msun$ \cite{GWCBC,BBHO1},
the previous works have only considered low-mass systems.
We therefore extend the study to high-mass systems up to $M=100\msun$ and compare our result with those of the previous works.
The purpose of this work is to examine the efficiency of a nonspinning bank 
for aligned-spin signals in a wide mass range.
To this end we investigate the range in spin  over which the nonspinning bank has fitting factors larger than 0.965
varying total mass and mass ratio of the signal.


\section{GW data analysis}


In signal processing, if a signal of known shape is buried in stationary Gaussian noise,
the matched filter can be the optimal method to identify the signal.
For the GWs emitted from merging  BBHs, since there exist various models that
can produce accurate full IMR waveforms, 
the matched filter can be employed in the BBH searches.
If a detector data stream $x(t)$ contains  stationary Gaussian noise $n(t)$ and  a GW signal $s(t)$,
the match between $x(t)$ and a template waveform $h(t)$ is determined by 
\be \label{eq.match}
\langle x | h \rangle = 4 {\rm Re} \int_{f_{\rm low}}^{\infty}  \frac{\tilde{x}(f)\tilde{h}^*(f)}{S_n(f)} df,
\ee
where the tilde denotes the Fourier transform of the time-domain waveform, $S_n(f)$ is the power spectral density (PSD) of the detector noise, and
$f_{\rm low}$ is the low frequency cutoff that depends on the shape of $S_n(f)$.
In this work, we consider a single detector configuration and use the zero-detuned, high-power noise PSD with $f_{\rm low}=10$ Hz \cite{apsd}.
Using the relation in Eq. (\ref{eq.match}), the signal-to-noise ratio $\rho$ (SNR) can be determined by 
\be
\rho=\langle s | \hat{h} \rangle,
\ee
where $\hat{h} \equiv h/ \langle h| h\rangle^{1/2}$ is the normalized template.
When the template waveform $h$  has the same shape as the signal waveform $s$, 
the matched filter gives the optimal SNR as
\be
\rho_{\rm opt}=\langle s|s \rangle^{1/2}.
\ee
If the template has a different shape, 
the SNR is reduced to
\be
\rho={\rm FF} \times \rho_{\rm opt},
\ee
where FF is the fitting factor defined as the best-match between a normalized signal and a set of normalized templates \cite{Apo95}.

To fully describe the wave function of an aligned-spin BBH system, we need 11 parameters except the eccentricity. 
Those are five extrinsic parameters (luminosity distance of the binary, two angles defining the sky position of the binary with respect to the detector, orbital inclination, and wave polarization), four intrinsic parameters (component masses and spins), the coalescence time $t_c$, and the coalescence phase $\phi_c$.
However, since the extrinsic parameters  only scale the wave amplitude,
and we work with the normalized wave function, we do not need to consider the extrinsic parameters in our analysis.
In addition, the inverse Fourier transform of the match can give the output for all possible coalescence times at once,
and  we can maximize the match over all possible coalescence phases by taking the absolute value of the complex-valued output (see \cite{All12} for more details).
Therefore, we need only the intrinsic parameters ($m_1, m_2, \chi_1, \chi_2$) in our analysis,
and those are the input parameters of PhenomD.

On the other hand, it is often more efficient to treat the effect of aligned-spins with a single spin parameter rather than the two component spins
because the two spins are strongly correlated \cite{San10,Aji11a, Pur13,Nie13,Pur16b}.
For this purpose, the spin effects in the phenomenological models are parametrized by an effective spin $\chi$:
\be \label{eq.effective spin}
\chi \equiv \frac{m_1 \chi_1+ m_2 \chi_2}{M}.
\ee
The value of $\chi$ can be determined simply by choosing  $\chi_1=\chi_2=\chi$ in the PhenomD wave function\footnote{
PhenomD is parametrized by a normalized reduced effective spin $\hat{\chi}$ \cite{Kha16}, but we can have $\hat{\chi}=\chi$ by choosing $\chi_1=\chi_2$.}.
Thus, our signal waveform is given by $h_s=h(m_1, m_2, \chi)=h_{\rm PhenomD}(m_1, m_2, \chi, \chi)$, while
the nonspinning templates  are given by $h_t=h(m_1,m_2)=h_{\rm PhenomD}(m_1, m_2, 0, 0)$.

In this work, we define the overlap $P$ by the match between  the signal $\hat{h}_s$ and the template $\hat{h}_t$ maximized over $t_c$ and $\phi_c$:
\be\label{eq.overlap}
P =  \max_{t_c,\phi_c}\langle \hat{h}_s | \hat{h}_t \rangle. 
\ee
Thus, we can have $P=1$ if the signal and the template have the same shapes.
Changing the mass parameters  of the nonspinning templates, we calculate the two-dimensional overlap surface as
\be\label{eq.2d overlap}
P(\lambda) =  \max_{t_c,\phi_c}\langle \hat{h}_s(\lambda_0) | \hat{h}_t(\lambda) \rangle, 
\ee
where $\lambda_0$ denotes the true values of the mass and the spin of the signal,
and $\lambda$ denotes the mass parameters of the template.
Then, in our analysis the fitting factor corresponds to the maximum value in the overlap surface: 
\be\label{eq.FF}
{\rm FF} = \max_{ \lambda} P(\lambda).
\ee

On the other hand, in an actual search for BBHs the template waveforms are discretely placed in the bank, 
hence the fitting factor can be marginally reduced
depending on the template density. 
Thus, the effective fitting factor is obtained by
\be\label{eq.effective FF}
{\rm FF}_{\rm eff} = \max_{h_t \in {\rm bank}}P(\lambda).
\ee
Typically, when one chooses a waveform model for the search, the template bank is constructed densely enough such  that the  mismatch between the templates and
the signal does  not  exceed $3\%$ including the effect of the discreteness of the template spacing, i.e. $1-{\rm FF}_{\rm eff} \geq 0.97$ \cite{Aba12,Aas13}.
In this work, however, 
we want to remove the effect of discreteness on the fitting factor.
To do so,  we choose sufficiently fine
spacings in the template space defined in the $M_c - \eta$ plane  \cite{Cho15c,Cho15d}.\footnote{
In general, the overlap surface is obtained more efficiently in the parameter space consisting of the chirp mass ($M_c \equiv (m_1 m_2)^{3/5}/M^{1/5}$) and the symmetric mass ratio ($\eta \equiv m_1 m_2 /M^2$), so  we take into account the parameters $M_c, \eta$ instead of $m_1, m_2$ in the overlap calculations.} 
 For example, in order to obtain ${\rm FF}_{\rm eff}$ for one signal,  we repeat a grid search around $\lambda_0$ 
 until we find the crude location of the peak point in the overlap surface. 
 Next, we estimate the size of the contour $\bar{P} \equiv P/ P_{\rm max} = 0.995$, 
 where $P_{\rm max}$ is the maximum overlap value in that contour (if the recovered mass parameters are biased from $\lambda_0$, then $P_{\rm max} < 1$), hence 
$\bar{P}$ corresponds to the weighted overlap. 
Finally, we find (almost) the exact location of the peak point by performing a $31 \times 31$ grid search
in the region of $\bar{P} > 0.995$, and the overlap value at the peak point is regarded as  ${\rm FF}$.

Once a fitting factor is determined through the above procedure,
we can measure the systematic bias, which corresponds to the distance from the true value $\lambda_0$ to the recovered value $\lambda^{\rm rec}$:
\be\label{eq.bias}
b =   \lambda^{\rm rec} - \lambda_0.
\ee
Typically, the recovered parameters are systematically biased from the true parameters if the incomplete template waveforms are used.
In our analysis, the incompleteness of templates arises from neglecting the spin effect in the wave function.
As the efficiency of a template waveform model for the search is evaluated by the fitting factor,
its validity for the parameter estimation can be  examined by the systematic bias.


\section{Result}

We choose as our target signals  aligned-spin BBHs  in the parameter regions of $m_1, m_2 \geq 5 \msun \ (m_2 \leq m_1), \  M\leq 100 \msun$ and  $-0.4 \leq \chi \leq 0.4$. 
The signal waveforms are generated by using PhenomD with $\chi_1=\chi_2=\chi$.
We construct a template bank in the $M_c -\eta$ plane with nonspinning waveforms assuming $\chi_1=\chi_2=0$ in PhenomD.
The templates are assumed to be placed densely enough so that we can avoid the effect of the discreteness of the bank.
Using the nonspinning templates with an aligned-spin signal we calculate the overlap surface that includes the confidence region, 
and determine the fitting factor and the systematic bias for the signal.


\subsection{Fitting factor}
\begin{figure*}[t]
\begin{center}
\includegraphics[width=15cm]{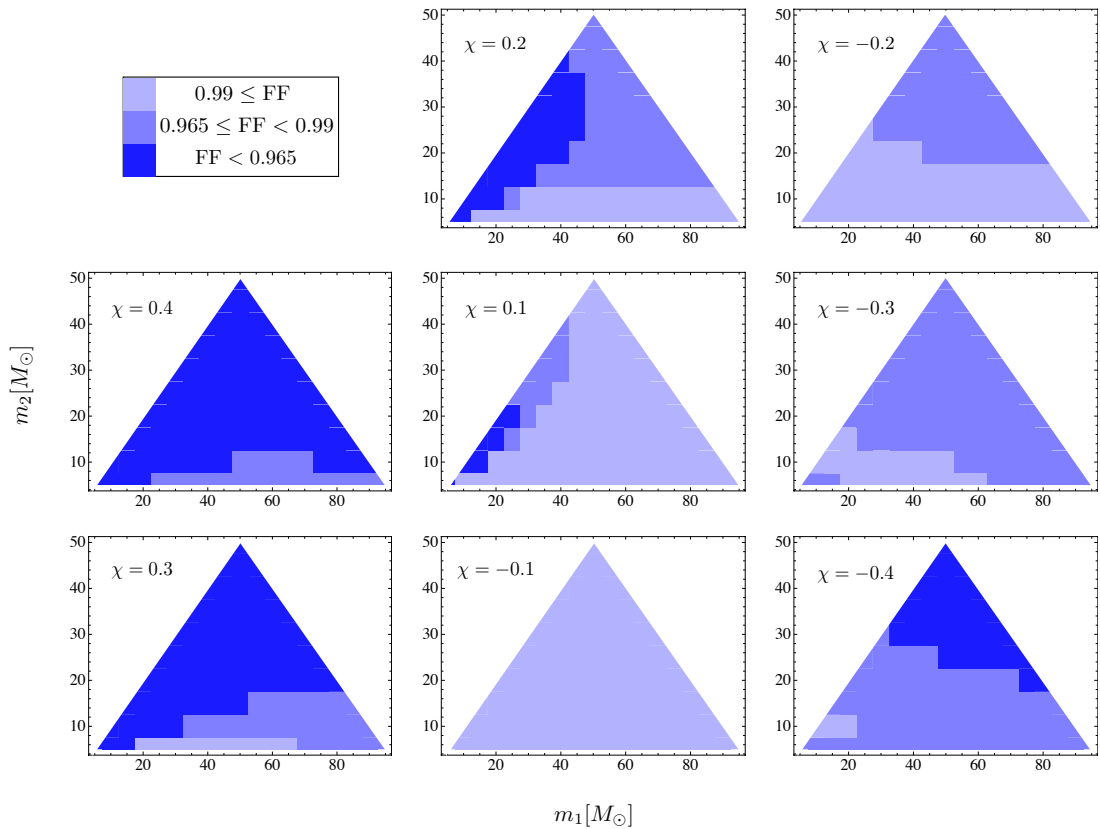}
\caption{\label{fig.ff-2d-1}Fitting factors obtained by using nonspinning templates for aligned-spin BBH signals.
The spin value of the signal is given in each panel. The signals with negative spins can have higher  fitting factors than those with positive spins}
\end{center}
\end{figure*}

\begin{figure}[t]
\begin{center}
\includegraphics[width=\columnwidth]{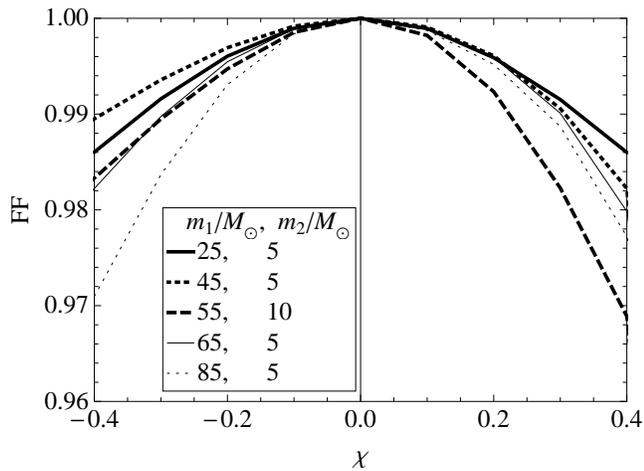}
\caption{\label{fig.ff-1d-1} Examples that have the fitting factors larger than 0.965 in the spin range of $-0.4 \leq \chi \leq 0.4$.}
\end{center}
\end{figure}

In Fig. \ref{fig.ff-2d-1}, we show the fitting factors for all of the BBH signals.
In each panel, the darkest region corresponds to the signals that cannot achieve the fitting factor exceeding a threshold of 0.965 beyond which a loss of detection rates does not exceed $\sim 10\%$.
We find that the signals with negative spins can have higher  fitting factors than those with positive spins.
If $\chi = 0.3$, only the highly asymmetric-mass signals can have the fitting factors exceeding the  threshold.
However, if $\chi = -0.3$, the fitting factors for all of the signals can be larger than the threshold,
and if  $\chi = -0.4$, about two third of the signals can have fitting factors exceeding the  threshold.
In particular,  if the signal has a small spin in the range of $-0.1\leq \chi \leq 0.1$, the fitting factor can be larger than 0.99 (the lightest region)
for all of the signals except those in the highly symmetric-mass region.
The range in spin over which all of the signals in our mass range have fitting factors exceeding 0.965 is very narrow, i.e.,
$-0.3 \leq \chi \leq 0$.
On the other hand, a few binaries can achieve ${\rm FF} \geq 0.965$ in our spin range,
and we show several examples in Fig. \ref{fig.ff-1d-1}.

\begin{figure*}[t]
\includegraphics[width=15cm]{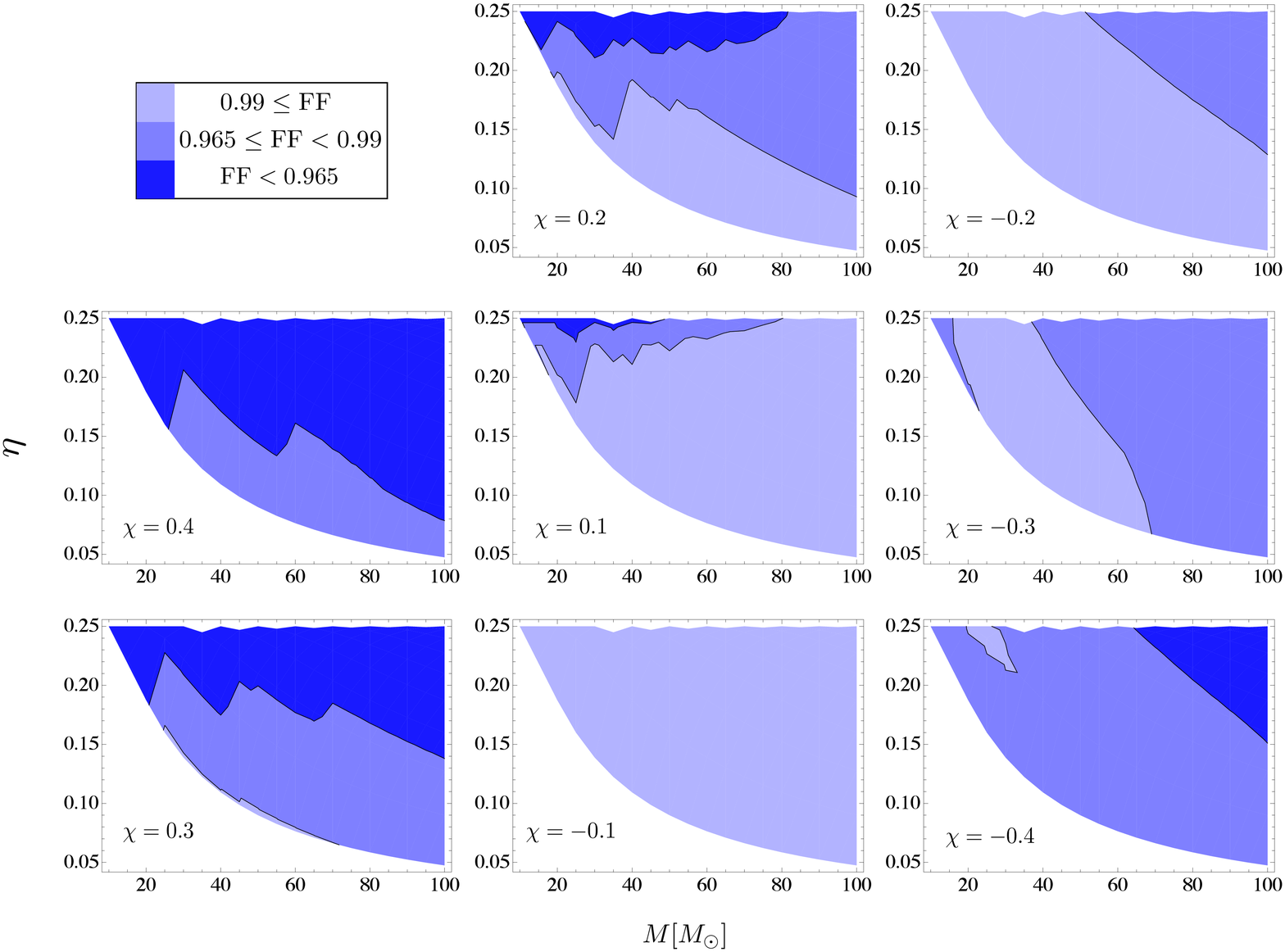}
\caption{\label{fig.ff-2d-2}The same fitting factors as in Fig. \ref{fig.ff-2d-1} but described in the $M-\eta$ plane.}
\end{figure*}

\begin{figure}

        \centering
        \subfloat[ \label{fig.ff-1d-2}]{%
                \includegraphics[width=0.5\textwidth]{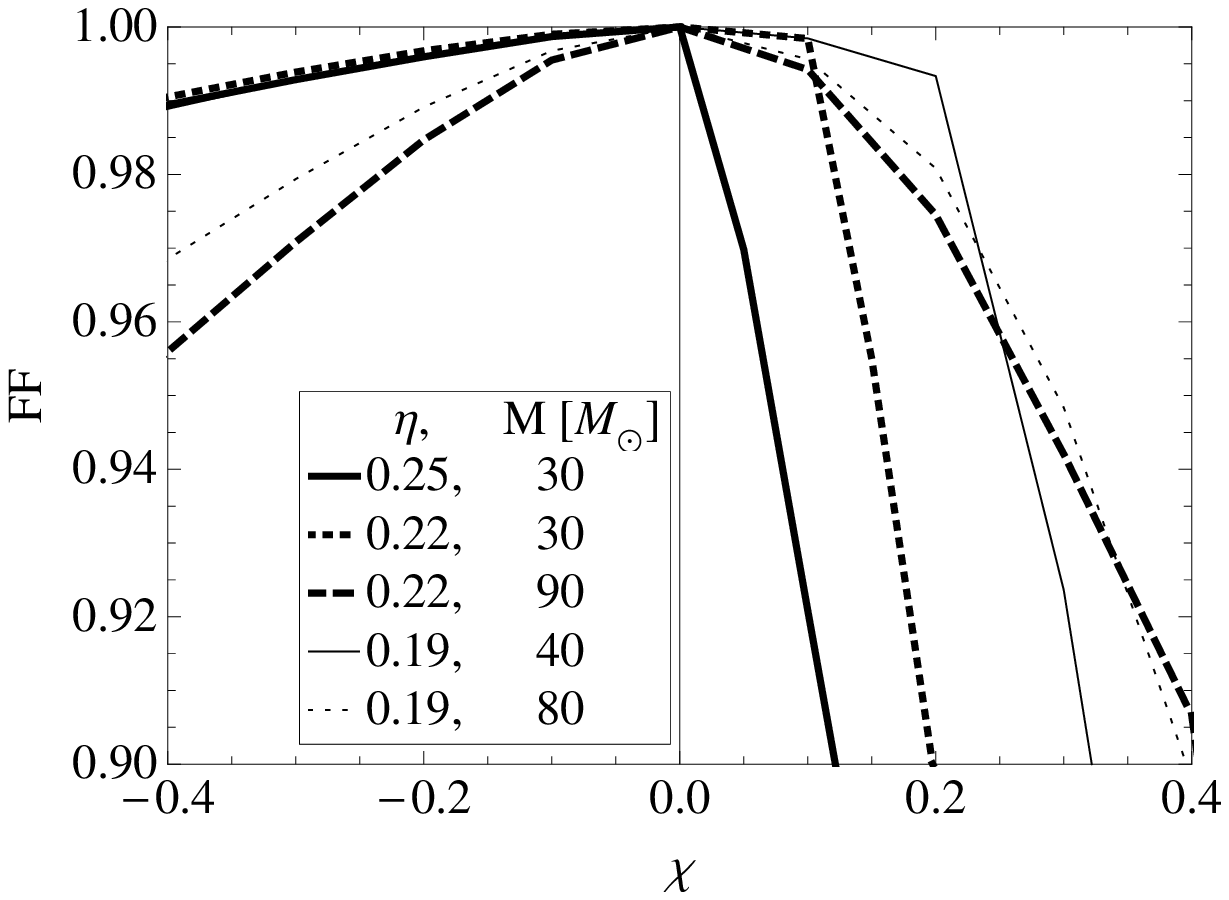}}

        \centering
        \subfloat[ \label{fig.bias-1d}]{%
                \includegraphics[width=0.5\textwidth]{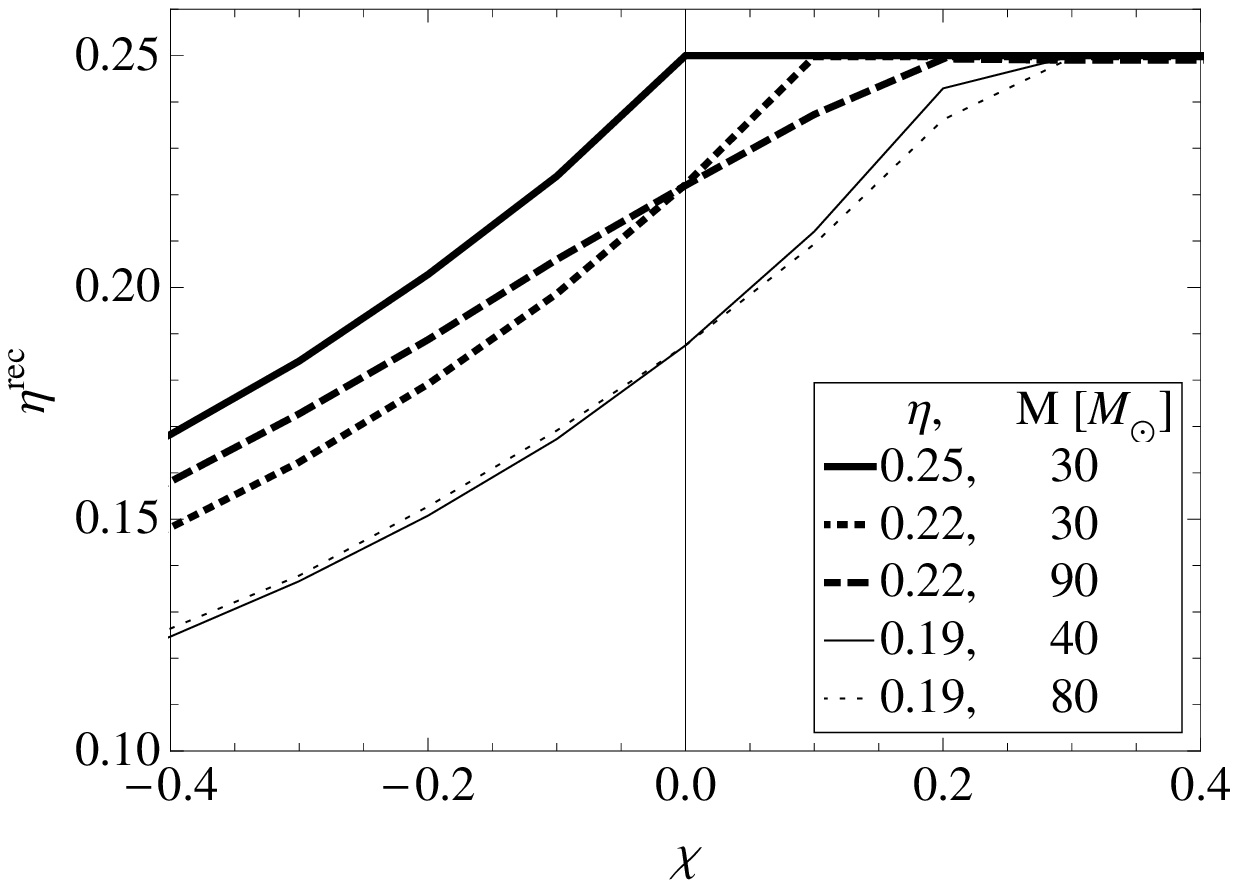}}

    \caption{\label{fig.ff-bias}Fitting factors and recovered  $\eta$ ($\eta^{\rm rec}$) for several binaries. When $\chi > 0$, the fitting factor suddenly falls off at a certain spin value (a).
    $\eta^{\rm rec}$ cannot exceed the physical boundary 0.25 (b).
The spin value at which $\eta^{\rm rec}$ reaches 0.25 is consistent with
the one at which the sudden fall-off of the fitting factor occurs.}
\end{figure}

In Fig. \ref{fig.ff-2d-2},  we also show the fitting factors in the $M-\eta$ plane using the same color scales as in  Fig. \ref{fig.ff-2d-1}.
In this figure, we can interpret the pattern of the fitting factors more easily.
In the region of a negative spin,  the fitting factor tends to decrease as the total mass or the symmetric mass ratio increases.
On the other hand, in the region of a positive spin, we can see a strong dependence of the fitting factor on the symmetric mass ratio.
In this case, the fitting factors in the symmetric-mass region rapidly decrease with increasing $\chi$, especially,
those with low masses can drop below the threshold even with the small spin of $\chi=0.1$.
In Fig. \ref{fig.ff-1d-2}, we show some examples that show highly asymmetric fitting factors between a positive and a negative spins.
We find that if $\chi > 0$, the fitting factor suddenly falls off at a certain spin value, and the falling rate tends to slacken for higher-masses.

Dal Canton \etal \cite{Dal14} showed that the sudden fall-off of the fitting factor is associated with the physical boundary of the template space.
For the (positively) aligned-spin signals, the parameter value of $\eta$ recovered by the nonspinning templates
increases as the spin of the signal increases. 
However, in the parameter space of ($M_c, \eta$), 
the physical value of $\eta$ should be restricted to the range of $0\leq \eta \leq 0.25$. 
Thus, the recovered value of $\eta$ cannot exceed 0.25 even though the signal has higher spins.
For example, Fig. \ref{fig.bias-1d} shows the recovered $\eta$ ($\eta^{\rm rec}$) as a function of $\chi$ for the same binaries as in Fig. \ref{fig.ff-1d-2}.
If the true value of $\eta$ is 0.25, $\eta^{\rm rec}$ is already at the boundary at $\chi=0$, hence always equal to 0.25 in the entire range of positive spins.
In Fig. \ref{fig.ff-bias}, we find that the spin value at which the $\eta^{\rm rec}$ reaches 0.25 is consistent with
the one at which the sudden fall-off of the fitting factor occurs.
On the other hand, the post-Newtonian waveforms are well
behaved for $0<\eta<1.0$ although the unphysical value of $\eta$ 
implies complex-valued masses.
Boyle \etal \cite{Boy09} showed that
the fitting factors for high-mass systems above $\sim 30 \msun$ can be significantly improved  
if $\eta$ is allowed to range over unphysical values.
However, such the unphysical masses are not permitted in the phenomenological models.

\subsection{Comparing with other works}

\begin{figure}[t]
\begin{center}
\includegraphics[width=\columnwidth]{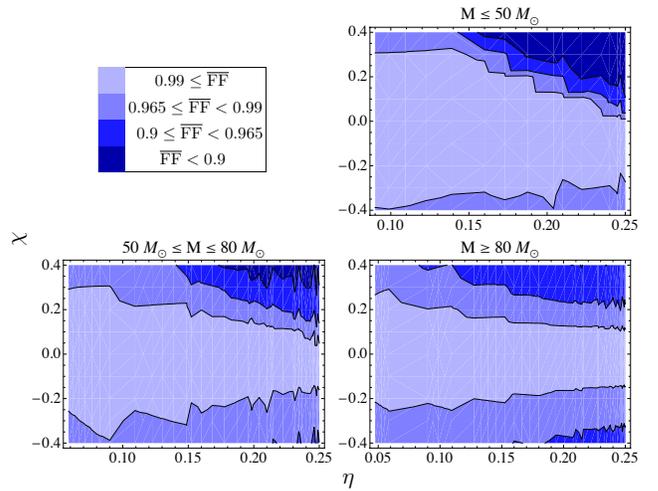}
\caption{\label{fig.ff-eta-chi} Mean fitting factors ($\overline{\rm FF}$) described in the $\eta-\chi$ plane for the low-mass (top), medium-mass (bottom left), and high-mass (bottom right) systems, respectively. The mean fitting factor is calculated by averaging over $M$.}
\end{center}
\end{figure}

\begin{figure}[t]
\begin{center}
\includegraphics[width=\columnwidth]{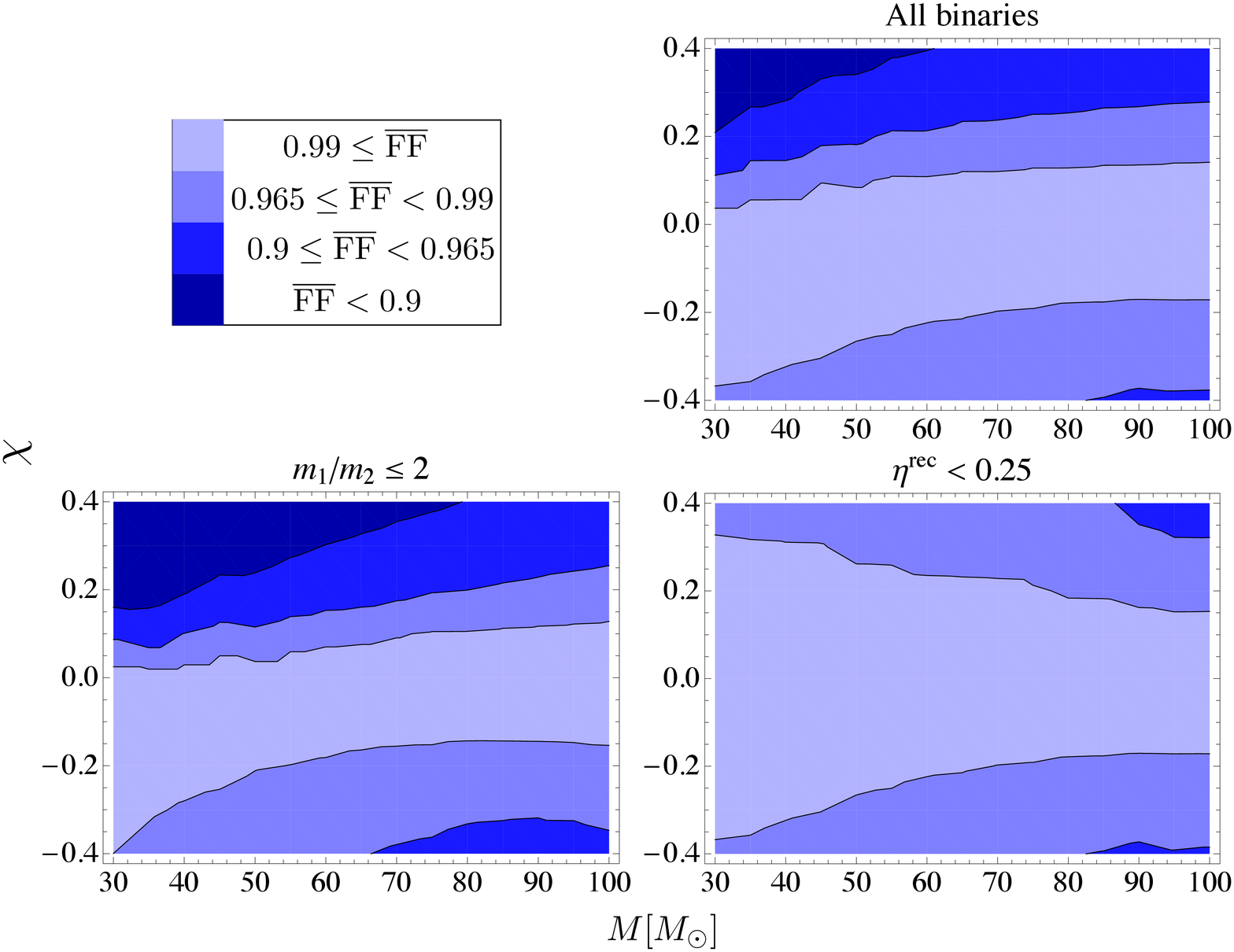}
\caption{\label{fig.ff-M-chi} Mean fitting factors ($\overline{\rm FF}$) described in the $M-\chi$ plane for all of the binaries with $M\geq 30\msun$ (top),
symmetric-mass binaries with $m_1/m_2 \geq 2$ (bottom left), and asymmetric-mass binaries for which $\eta_{\rm rec}<0.25$, respectively.
The mean fitting factor is calculated by averaging over $\eta$.}
\end{center}
\end{figure}

In Fig. \ref{fig.ff-eta-chi},  we represent the fitting factors in the $\eta-\chi$ plane in a different way.
We classify our binaries into low-mass ($M\leq 50 \msun$), medium-mass ($50 \msun \leq M \leq 80 \msun$), and high-mass ($80 \msun \leq M$) systems,
and calculate the mean fitting factors ($\overline{\rm FF}$) by averaging over $M$  for  each system.
Note that since we assume the minimum mass of $m_2$ to be $5 \msun$, the values of $\eta$ start from 0.09 (top),  $\sim 0.06$ (bottom left), and $\sim 0.05$ (bottom right), respectively.
The range of $\chi$, in which $\overline{\rm FF} \geq 0.99$, becomes smaller as $\eta$ increases.
For the low-mass systems, the fitting factor curves in the region of a positive spin rapidly drop to zero.
Dal Canton \etal \cite{Dal14} also described the fitting factors in the same manner for  BH-NS binaries with masses of $M\leq 18 \msun$ (see Fig. 8 therein),
and our result for the low-mass system shows the pattern of fitting factor similar to their result in the region of a positive spin.
However,  in the region of a negative spin, they had poor fitting factors, and
they pointed out that this is because the minimum NS mass in the template bank is limited to $1\msun$.
In particular, we find that the overall area with high fitting factors is narrower for  higher-mass systems.
That means the nonspinning bank has worse search efficiency for higher-mass systems. 

We also describe the fitting factors in the $M-\chi$ plane in  Fig. \ref{fig.ff-M-chi} and compare those with the result of Privitera \etal \cite{Pri14}.
While Privitera \etal  considered low-mass BBHs in the range of  $M\leq 35 \msun$ with the initial LIGO PSD  \cite{ipsd}  assuming $f_{\rm low}=40$ Hz, 
we take into account the higher-mass binaries in the range of $M\geq 30 \msun$\footnote{
Since we have only few  samples in the range of $M < 30 \msun$, we do not include the results for those binaries in this figure.}
 with the Advanced LIGO PSD  \cite{apsd} assuming $f_{\rm low}=10$ Hz.
Therefore,  our result cannot be directly compared with their result.
However, we find that the overall pattern of the fitting factors in our result is similar to
the result of \cite{Pri14} (see, Fig. 1 (a) therein).
The top panel in Fig. \ref{fig.ff-M-chi} shows very asymmetric fitting factors between the regions of a positive and a negative spins.
For positive spins, the fitting factor contours 
gradually increase as the total mass increases, and this is roughly consistent with the result of \cite{Pri14}.
On the contrary, for negative spins, the range of $\chi$, in which the signals have high fitting factors, 
is much larger than  the case for positive spins in the low-mass region, but that becomes smaller as the total mass increases.
We already showed that the discrepancy between the  two spin regions is caused by the physical boundary of the template space.
To see this concretely, we select only the symmetric-mass binaries with $m_1/m_2 \leq 2$ and show their results in the bottom left panel in Fig. \ref{fig.ff-M-chi}.
We find that the discrepancy is more pronounced compared to the result of the top panel. 
We also choose the asymmetric-mass binaries for which $\eta^{\rm rec}$ does not reach the physical boundary, i.e., $\eta_{\rm rec}<0.25$,
and show their results in the bottom right panel.
As expected, we can see nearly symmetric  fitting factors between  the regions of a positive and a negative spins.
Especially, in this case, most of the binaries can have mean fitting factors greater than $0.965$.
That means, in the nonspinning template search for aligned-spin BBH signals, 
most of the signals,  that have the masses of $M\leq 100\msun$ and the spins of $-0.4 \leq \chi \leq 0.4$,
have high fitting factors exceeding the threshold 0.965
if only the binary has the asymmetric masses such that $\eta^{\rm rec}$ does not reach 0.25.


\subsection{Systematic bias of the recovered parameter}
Once a detection is made  in the search pipeline,
the parameter estimation pipeline conducts post-processing with
the data stream, that contains the GW signal.
The purpose of the parameter estimation analysis is
to extract the parameters of a signal with high accuracy \cite{Aas13}.
The results of the parameter estimation are given by the posterior probability density functions for the parameters \cite{GW1PE1, Aas13b, Vei15}.
Usually the posterior probability distribution is sampled by the Markov-chain Monte Carlo or nested sampling methods \cite{Vei15}.
However, these algorithms are computationally intensive.
In the high SNR limit,
the Fisher matrix  method can be used to approximate the statistical error in the parameter estimation \cite{Cho13,Cho14,Cho15a,Cho15b,Cho15e} (for more details refer to \cite{Val08} and references therein).

\begin{figure}[t]
\begin{center}
\includegraphics[width=\columnwidth]{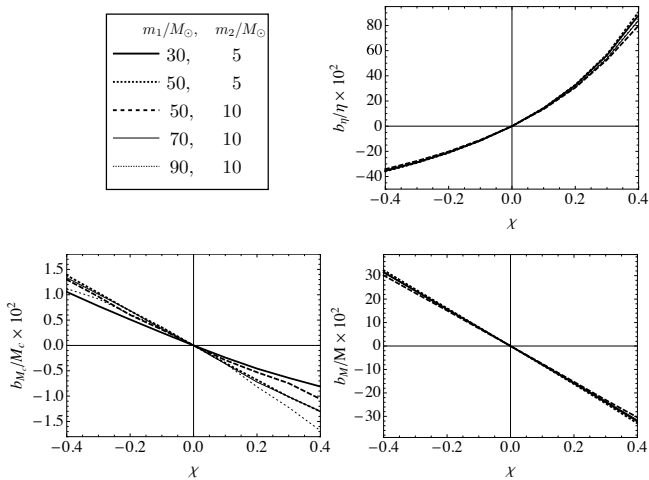}
\caption{\label{fig.bias-1d-all}Systematic bias ($b_{\lambda}/\lambda$) as a function of $\chi$ for $\eta$ (top), $M_c$ (bottom left), and $M$ (bottom right), respectively.}
\end{center}
\end{figure}

On the other hand, in the search, parameters of a signal can also be
inferred  from the identified template parameters, but the recovered parameters 
can be significantly biased from the true parameters.
In this subsection,
we show how much the recovered parameter is biased   
depending on the spin of the signal.
In Fig. \ref{fig.bias-1d-all}, we show the fractional bias ($b_{\lambda}/\lambda$) as a function of $\chi$.
Here, as concrete examples we select several asymmetric-mass binaries  that satisfy $\eta^{\rm rec} < 0.25$.
In the top panel, as $\chi$ increases  the bias for $\eta$ also increases, and the dependence of the bias on $\chi$ is stronger for a positive spin than a negative spin.
On the contrary, in the bottom left panel, the bias for $M_c$ decreases with increasing $\chi$, and that exhibits a similar dependence on $\chi$ between a positive and a negative spins.
The biases incorporated in the two mass parameters can be well understood by describing those in terms of a total mass.
When the spin is positively aligned with the orbital angular momentum, the spin-orbit coupling makes the binary's phase evolution slightly slower, hence delays  the onset of the plunge phase, as compared to its nonspinning counterpart \cite{Cam06}.
On the contrary, in the antialigned case, the phase evolution becomes slightly faster, and  the plunge is hastened.
Consequently, for a given starting GW frequency,  a positively (negatively) aligned-spin increases (decreases) the
length of the waveform, as compared to the nonspinning case. 
Therefore,  positively (negatively) spinning systems
can be recovered by lower (higher) mass nonspinning templates.
We clearly describe this in the bottom right panel, showing the bias for the parameter $M$ as a function of $\chi$.
Interestingly, we find that  the systematic bias for  $M$ almost linearly depends on $\chi$ in our spin range.
In addition, all of the results seem to have similar fractional biases ($b_M/M$) for a given $\chi$ even though their masses are very different.

\begin{figure}[t]
        \includegraphics[width=\columnwidth]{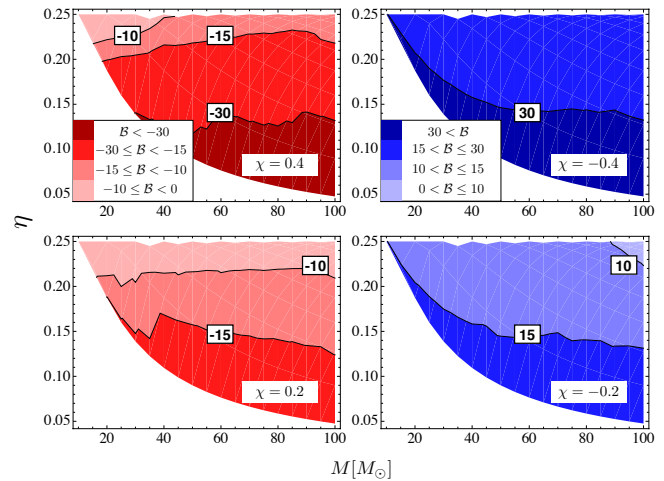}
    \caption{\label{fig.bias-M-2d} Fractional biases for the parameter $M$ (${\cal B} \equiv b_M/M \times 10^2$). The biases are similar between a positive (red) and a negative (blue) spins in the asymmetric-mass region ($\eta \lesssim 0.15$). The similarity between the contours ${\cal B}=15 \ (-15)$ and  $30 \ (-30)$ indicates a linear relation between ${\cal B}$ and $\chi$.}
\end{figure}

In Fig. \ref{fig.bias-M-2d}, we show the fractional biases (${\cal B} \equiv b_M/M$) for the signals with the spins of $\chi=-0.4, -0.2, 0.2,$ and $0.4$.
The red color indicates a negative bias while the blue color indicates a positive bias.
We find that the magnitudes of biases are similar between the red and the blue 
in the asymmetric-mass region ($\eta \lesssim 0.15$), while those are smaller for the positive spins in the symmetric-mass region ($\eta \gtrsim 0.15$).
As expected, the difference in the symmetric-mass region is due to the fact that for the positive spins
$\eta^{\rm rec}$ is restricted by the physical boundary, and thereby  the corresponding $M^{\rm rec}$ has smaller biases.
We also find that the contours ${\cal B}=30, -30$ in the top panels are consistent with the contours ${\cal B}=15, -15$ in the bottom panels,
and this indicates a linear relation between ${\cal B}$ and $\chi$.
Finally, we find that in the asymmetric-mass region all of the fractional biases are comparable  for a given $\chi$
independently of the total mass. For example, we have $15 \ (30)  \lesssim {\cal B} \lesssim 17 \ (35)$ for 
$\chi=-0.2 \ (-0.4)$.

\section{Summary and discussion}\label{sec5}
We investigated the efficiency of nonspinning templates in GW searches for aligned-spin BBHs.
We considered the signals with moderately small spins in the range of $-0.4 \leq \chi \leq 0.4$.
We employed as our waveform model PhenomD,
and we set the spins to zero for the nonspinning waveforms.
Using the nonspinning templates, 
we calculated the fitting factors of the aligned-spin BBH signals in a wide mass range up to $\sim 100\msun$.
The results are summarized in Figs. \ref{fig.ff-2d-1} and \ref{fig.ff-2d-2} in the $m_1-m_2$ plane and $M-\eta$ plane, respectively.
The signals with negative spins can have higher fitting factors than those with positive spins.
If $\chi = 0.3$, only the highly asymmetric-mass signals can have the fitting factors exceeding the threshold 0.965. 
However, if $\chi = -0.3$, the fitting factors for all of the signals can be larger than the threshold.
The discrepancy between the regions of a positive and a negative spin is due to the fact that
the template parameter space is physically restricted to $\eta \leq 0.25$ so that the recovered value of $\eta$ ($\eta^{\rm rec}$) cannot exceed 0.25. 
We demonstrated this by choosing the asymmetric-mass binaries that satisfy $\eta^{\rm rec}< 0.25$,
and showing the nearly symmetric fitting factors for those binaries between the two regions.
We classified our binaries into low-mass, medium-mass, and high-mass systems and calculated the mean fitting factor by averaging over $M$ in the $\eta-\chi$ plane,
and found that the overall area with high fitting factors is narrower for higher-mass systems.
The mass parameters recovered by the nonspinning templates are significantly biased 
from the true parameters of the aligned-spin signals.

In this work, we revisited the issues on the effectualness of nonspinning templates in aligned-spin BBH searches
that were addressed in several works for low-mass BBHs.    
We obtained a similar result to those of the previous works 
and found that the nonspinning bank has worse search efficiency for higher-mass systems.
Overall, we obtained a very narrow range in spin ($-0.3\leq\chi\leq0$) over which the nonspinning bank
has fitting factors exceeding 0.965 for all of the aligned-spin signals in our mass range.
Moreover, the fitting factors given in this work should be a bit lowered if the discreteness of template spacing
is considered in our analysis. 
Therefore, our study demonstrates the ineffectualness of the nonspinning bank
and emphasizes the necessity of aligned-spin templates in the current
Advance LIGO searches for aligned-spin BBHs.


%

\section*{ACKNOWLEDGMENTS}
The author would like to thank Stephen Privitera, Tito Dal Canton, and Alex Nielsen for helpful comments. This work was supported by the National Research Foundation of Korea (NRF) grant funded by the Korea government (Ministry of Science, ICT \& Future Planning) (No. 2016R1C1B2010064).
This work used the computing resources at the KISTI Global Science Experimental Data Hub Center (GSDC).
%
%
%


\begin{thebibliography}{9}
\bibitem{GW1} B. P. Abbott \etal (LIGO Scientific Collaboration and Virgo Collaboration), {\prl}  {\bf 116}, 061102 (2016).
\bibitem{GW2} B. P. Abbott \etal (LIGO Scientific Collaboration and Virgo Collaboration), {\prl}  {\bf 116}, 241103 (2016).
\bibitem{ALIGO} J. Abadie \etal (LIGO Scientific Collaboration), Classical Quantum Gravity {\bf 32}, 074001 (2015).
\bibitem{AVirgo} F. Acernese \etal, Classical Quantum Gravity {\bf 32}, 024001 (2015).
\bibitem{KAGRA} Y. Aso \etal (The KAGRA Collaboration), {\prd} {\bf 88}, 043007 (2013).
\bibitem{Aba10} J. Abadie \etal (LIGO Scientific Collaboration and Virgo Collaboration), Classical Quantum Gravity  {\bf 27}, 173001 (2010).
\bibitem{Dom14} M. Dominik, E. Berti, R. O'Shaughnessy, I. Mandel, K. Belczynski, C. Fryer, D. Holz, T. Bulik and F. Pannarale, {\apj} {\bf 806}, 263 (2015).
\bibitem{Abb16} B. P. Abbott \etal (LIGO Scientific Collaboration and Virgo Collaboration), Astrophys. J. Lett.  {\bf 833}, L1 (2016).
\bibitem{Abb16b} B. P. Abbott \etal (LIGO Scientific Collaboration and Virgo Collaboration), Phys. Rev. X {\bf 6}, 041015 (2016).
\bibitem{GW1PE1} B. P. Abbott \etal (LIGO Scientific Collaboration and Virgo Collaboration), {\prl}  {\bf 116}, 241102 (2016).
\bibitem{GW1PE2} B. P. Abbott \etal (LIGO Scientific Collaboration and Virgo Collaboration), Phys. Rev. X {\bf 6}, 041014 (2016).
\bibitem{Nie16} A. B. Nielsen, J. Phys.: Conf. Ser.  {\bf 716}, 012002 (2016).

\bibitem{Pur14} M. P{\"u}rrer, Classical Quantum Gravity {\bf 31}, 195010 (2014).
\bibitem{Pur16} M. P{\"u}rrer, {\prd} {\bf 93}, 064041 (2016).
\bibitem{Tar12} A.  Taracchini \etal, {\prd} {\bf 86}, 024011 (2012).
\bibitem{Tar14} A. Taracchini \etal, {\prd} {\bf89}, 061502 (2014).



\bibitem{Aji07a} P. Ajith  \etal,  Classical Quantum Gravity {\bf 24}, S689 (2007).
\bibitem{Aji08a} P. Ajith \etal,  {\prd} {\bf 77}, 104017 (2008); {\bf 79}, 129901(E) (2009).
\bibitem{Aji08b} P. Ajith,   Classical Quantum Gravity {\bf 25}, 114033 (2008).
\bibitem{Aji11b} P. Ajith  \etal,  {\prl} {\bf 106}, 241101 (2011).
\bibitem{San10} L. Santamaria  \etal,  {\prd} {\bf 82}, 064016 (2010).
\bibitem{Han14} M. Hannam  \etal,  {\prl} {\bf 113}, 151101 (2014).
\bibitem{Kha16} S. Khan \etal, {\prd} {\bf 93}, 044007 (2016).
\bibitem{Kum16} P. Kumar \etal, {\prd} {\bf 93}, 104050 (2016).
\bibitem{Apo95}  T. A. Apostolatos, {\prd} {\bf 52}, 605  (1995).
\bibitem{Aji11a}  P. Ajith {\prd} {\bf84}, 084037 (2011). 
\bibitem{Dal14}  T. Dal Canton \etal, {\prd} {\bf90}, 082004 (2014).
\bibitem{Pri14}  S. Privitera \etal, {\prd} {\bf89}, 024003 (2014).
\bibitem{Cap16} C. Capano, I. Harry, S. Privitera, and A. Buonanno, {\prd} {\bf93}, 124007 (2016).
\bibitem{Har16} I. Harry, S. Privitera, A. Boh\'e, and A. Buonanno, {\prd} {\bf94}, 024012 (2016).

\bibitem{Aji14} P. Ajith, N. Fotopoulos, S. Privitera, A. Neunzert, N. Mazumder, and A. J. Weinstein, {\prd} {\bf 89}, 084041 (2014).
\bibitem{Har14} I. W. Harry, A. H. Nitz, D. A. Brown, A. P. Lundgren, E. Ochsner, and D. Keppel, {\prd} {\bf 89}, 024010 (2014).
\bibitem{Dal15}  T. Dal Canton, A. P. Lundgren, and A. B. Nielsen, {\prd} {\bf91}, 062010 (2015).



\bibitem{GWCBC} B. P. Abbott \etal (LIGO Scientific Collaboration and Virgo Collaboration), {\prd}  {\bf 93}, 122003 (2016).
\bibitem{BBHO1} B. P. Abbott \etal (LIGO Scientific Collaboration and Virgo Collaboration), Phys. Rev. X {\bf 6}, 041015 (2016).

\bibitem{apsd}{\it Advanced LIGO anticipated sensitivity curves}, https://dcc.ligo.org/LIGO-T0900288/public.
\bibitem{All12} B. Allen, W. G. Anderson,  P. R. Brady, D. A. Brown, and  J. D. E. Creighton,  {\prd} {\bf 85}, 122006 (2012).





\bibitem{Pur13} M. P{\"u}rrer, M. Hannam, P. Ajith, and S. Husa, {\prd} {\bf 88}, 064007 (2013).
\bibitem{Nie13} A. B. Nielsen,  Classical Quantum Gravity {\bf 30}, 075023 (2013).
\bibitem{Pur16b} M. P{\"u}rrer, M. Hannam,  and F. Ohme, {\prd} {\bf 93}, 084042 (2016).


\bibitem{Aba12} J. Abadie \etal (LIGO Collaboration, Virgo Collaboration), {\prd} {\bf 85}, 082002 (2012).
\bibitem{Aas13}   J.  Aasi  \etal (LIGO Scientific Collaboration, Virgo Collaboration), {\prd} {\bf 87}, 022002 (2013).
\bibitem{Cho15c} H.-S. Cho,  Classical Quantum Gravity {\bf 32}, 215023 (2015).
\bibitem{Cho15d} H.-S. Cho,  Classical Quantum Gravity {\bf 32}, 235007 (2015).
\bibitem{Boy09}  M. Boyle,  D. A. Brown, and  L. Pekowsky, Classical Quantum Gravity {\bf 26}, 114006 (2009).
\bibitem{ipsd} A. Lazzarini and R. Weiss, Report No. LIGOE950018-02-E, 1996.

\bibitem{Aas13b}   J. Aasi \etal (LIGO Scientific Collaboration, Virgo Collaboration), {\prd} {\bf 88}, 062001 (2013).
\bibitem{Vei15} J. Veitch \etal, {\prd}  {\bf 91}, 042003 (2015).




\bibitem{Cho13} H.-S. Cho, E. Ochsner, R. O'Shaughnessy, C. Kim, and C.-H. Lee,   {\prd} {\bf 87}, 024004 (2013).
\bibitem{Cho14} H.-S. Cho and C.-H. Lee,   Classical Quantum Gravity {\bf 31}, 235009 (2014).
\bibitem{Cho15a} H.-S. Cho,   {J. Korean Phys. Soc.} {\bf 66}, 1637 (2015).
\bibitem{Cho15b} H.-S. Cho,   {J. Korean Phys. Soc.} {\bf 67}, 960 (2015).
\bibitem{Cho15e} H.-S. Cho,  {J. Korean Phys. Soc.} {\bf 69}, 884 (2016).
\bibitem{Val08} M. Vallisneri, {\prd} {\bf 77}, 042001 (2008).


\bibitem{Cam06} M. Campanelli, C. O. Lousto, and Y. Zlochower, {\prd} {\bf 74}, 041501 (2006).






\end{thebibliography}
\end{document}